\documentclass{optica-article}

\journal{opticajournal} 

\articletype{Research Article}

\usepackage{lineno}
\usepackage{siunitx}
\usepackage{setspace} 

\begin{document}

\title{Polarization Purity and Dispersion Characteristics of Nested Antiresonant Nodeless Hollow-Core Optical Fiber at Near- and Short-wave-IR Wavelengths for Quantum Communications}

\author{
    Ivi Afxenti,\authormark{1}
    Lijun Yu,\authormark{1}
    Taylor Shields,\authormark{1}
    Daniele Faccio,\authormark{2}
    Thomas Bradley,\authormark{3,4}
    Lucia Caspani,\authormark{5}
    Matteo Clerici,\authormark{1,6} and
    Adetunmise C. Dada\authormark{2,*}
}

\address{
    \authormark{1}James Watt School of Engineering, University of Glasgow, Glasgow G12 8QQ, United Kingdom\\
    \authormark{2}School of Physics and Astronomy, University of Glasgow, Glasgow G12 8QQ, United Kingdom\\
    \authormark{3} Optoelectronics Research Centre, Zepler Institute, University of Southampton, SO17 1BJ, United Kingdom\\
     \authormark{4} Now at High-Capacity Optical Transmission Laboratory, Eindhoven University of Technology, 5600 MB, Netherlands\\
    \authormark{5}Institute of Photonics, Department of Physics, University of Strathclyde, Glasgow G1 1RD, United Kingdom\\
    \authormark{6}Dipartimento di Scienza e Alta Tecnologia, Università degli Studi dell'Insubria, Via Valleggio 11, 22100, Como, Italy
}

\email{\authormark{*}adetunmise.dada@glasgow.ac.uk} 


\begin{abstract*} 
Advancements in quantum  communication and sensing require improved optical transmission that ensures excellent state purity and reduced losses. While free-space optical communication is often preferred, its use becomes challenging over long distances due to beam divergence, atmospheric absorption, scattering, and turbulence, among other factors. In the case of polarization encoding, traditional silica-core optical fibers, though commonly used, struggle with maintaining state purity due to stress-induced birefringence. Hollow core fibers, and in particular nested antiresonant nodeless fibers (NANF), have recently been shown to possess unparalleled polarization purity with minimal birefringence in the telecom wavelength range using continuous-wave (CW) laser light. Here, we investigate a 1-km NANF designed for wavelengths up to the 2-$\mu$m waveband. Our results show a polarization extinction ratio between \textasciitilde-30 dB and \textasciitilde-70 dB across the 1520 to 1620 nm range in CW operation, peaking at \textasciitilde-60 dB at the 2-$\mu$m design wavelength. Our study also includes the pulsed regime, providing insights beyond previous CW studies, e.g., on the propagation of broadband quantum states of light in NANF at 2 $\mu$m, and corresponding extinction-ratio-limited quantum bit error rates (QBER) for prepare-measure and entanglement-based quantum key distribution (QKD) protocols. Our findings highlight the potential of these fibers in emerging applications such as QKD, pointing towards a new standard in optical quantum technologies.

\end{abstract*}

\section{Introduction}
Experimental realizations of quantum key distribution (QKD), in the second half of last century, established quantum communication as an important aspect of quantum technologies~\cite{bennett1989}. Since then, entanglement in various photonic degrees of freedom has been used to experimentally demonstrate various protocols, including coin tossing \cite{berlin2011}, and quantum money  \cite{schiansky2023} using time-bin and polarization encoding, respectively. Successful implementation of such quantum-communication protocols over long distances require optical networks with minimal losses. Consequently, existing telecommunication fiber infrastructures have been employed to transmit polarization qubits in the C-band due to their low absorption and propagation losses \cite{agrawal2021, senior2009}.

Although standard silica-core single-mode fibers (SMF) have seen significant advances in recent decades,  current fiber-networks face capacity limitations due to increasing demand for lower latency and higher data rates per wavelength band  \cite{gunning2019,richardson2010}. Hollow-core fibers (HCF) promise to address those challenges since they have the potential of outperforming standard silica fibers in terms of losses, while manifesting lower latencies and operating over larger bandwidths \cite{sakr2019,hong2021}. They have the advantage of confining the propagating light in air or other gases, which results in reduced nonlinear effects and offers a platform to study gas-light interactions \cite{ouzounov2003,yu2020}. In contrast to solid-core fibers, HCFs can be better customized for the chosen wavelength range of operation, which results in minimized propagation losses and dispersion, while offering robust single mode guidance \cite{sakr2020}. 	Recent pioneering work has begun to  explore the use of HCFs for QKD, highlighting their potential for reduced latency and crosstalk, environmental stability for long-distance applications~\cite{bt2021trial,zdnet2021quantum}, and the coexistence of classical and quantum signals \cite{zhang2024enhanced,li2022coexistence,oxford2021coexistence,Nasti:22}. 
 
 A newly emergent HCF technology, known as Nested Antiresonant Nodeless Fiber (NANF), demonstrates a significant reduction in confinement losses. This improvement is achieved by increasing the number of coherent air-glass reflections in the radial direction \cite{poletti2014}. Recent studies have reported exceptional polarization purity of up to -70dB/km at 1550nm in NANF, providing a leap in performance for polarization-sensitive applications~\cite{taranta2020}. Such polarization purity, coupled with the reported 0.28 dB/km and more recently 0.11 dB/km attenuation in the C and L bands \cite{jasion2020,Chen:24}, could facilitate the distribution of polarization-entangled states within a spectral region previously inaccessible---specifically, at 2 $\mu$m. This region is being explored as a means to expand the capacity of existing fiber networks. Consequently, it could help circumvent a potential `capacity crunch' by leveraging wavelength bands beyond the conventional telecom spectrum. The promise of HCF technology is also evidenced by a recent demonstration of the distribution of telecom time-bin entangled photons through a 7.7 km antiresonant HCF~\cite{antesberger2023distribution}.

The exploration of the 2 $\mu$m spectral window indeed goes beyond conventional applications in communication. Notably, both squeezed light and entangled photons have been demonstrated at $\sim$$2~\mu$m~\cite{mansell2018observation,Prabhakar2020two,Dada2021Near}, marking a significant leap in the realm of high-sensitivity metrology. These advancements are particularly crucial for applications requiring high-sensitivity detection. The LIGO-Voyager project exemplifies this advancement, leveraging the enhanced sensitivity offered by the 2 $\mu$m spectral window to refine its detection capabilities \cite{Ganapathy2023}. Furthermore, the intrinsic properties of silicon at this wavelength, including diminished linear and nonlinear losses, have catalyzed a paradigm shift towards the 2 $\mu$m region in the field of integrated quantum photonics \cite{Rosenfeld2020Mid,Shen2022}. This transition not only underscores the versatility of the 2 $\mu$m spectral window but also paves the way for groundbreaking innovations across various scientific and technological domains.

Here, we demonstrate and analyze the propagation of broadband states of light at 2 $\mu$m over a 1-km-long hollow-core NANF. This exploration addresses some of the technical challenges in expanding the operational wavebands for quantum communications, with our findings showcasing the potential of NANF technology. 

\section{Methods}
\begin{figure}[t!]
    \centering
\includegraphics[width=0.7\textwidth]{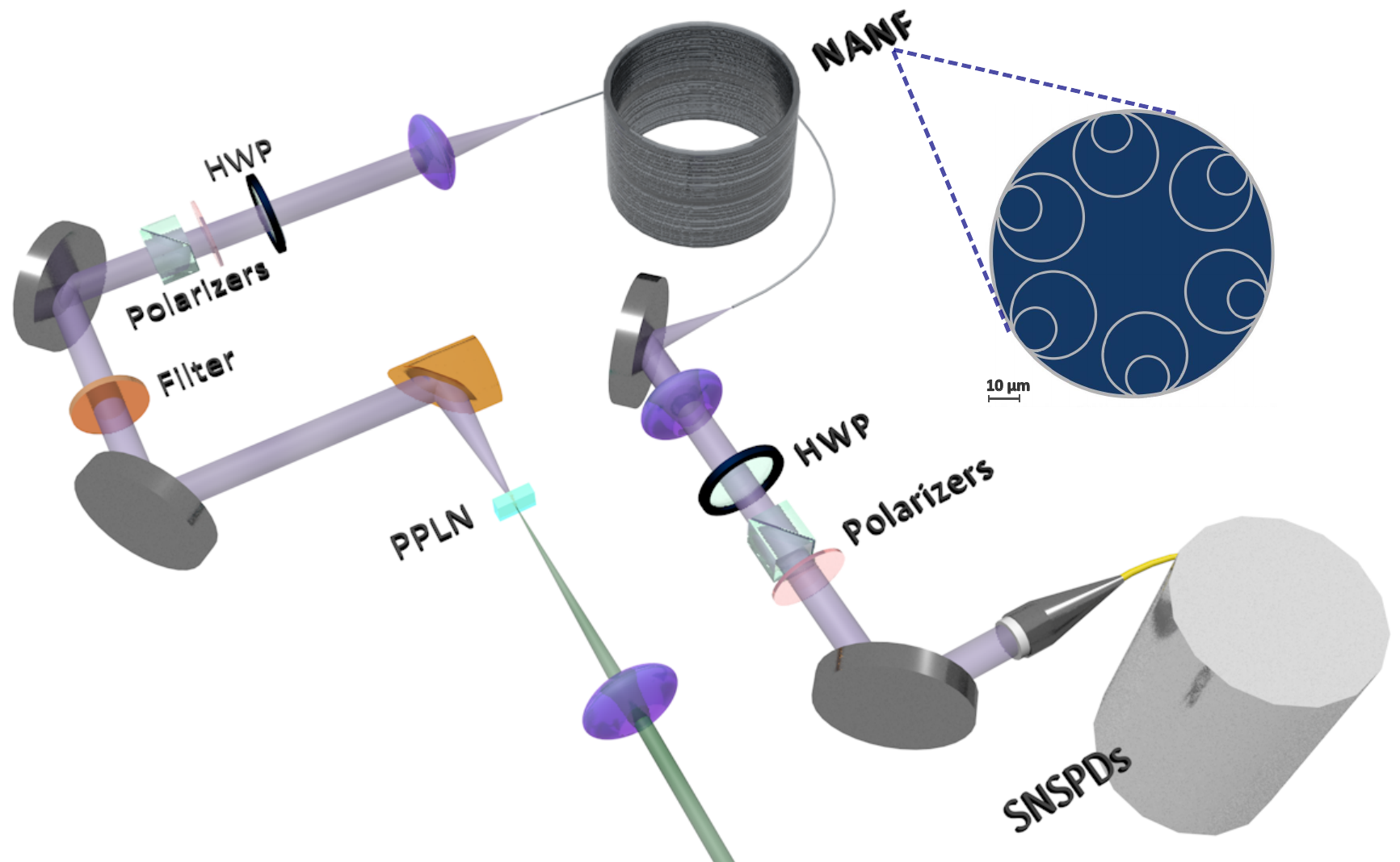}
    \caption{Experimental setup for the measurement of polarization extinction ratio and dispersion of photons propagating through a 1.1-km-long reel of nested antiresonant nodeless hollow-core fiber (NANF). The inset shows the cross-sectional structure of the fiber.}
    \label{fig:experimental_setup}
\end{figure}

To characterize the attenuation across the 1-km-long NANF, continuous wave (CW) fiber laser sources at 1560 nm and 2~$\mu$m wavelengths were used. The fiber, coiled into a spool (wound with a winding machine on a drum at a tension of 12 $g$)  with a diameter of 30~cm,  features a core diameter of 37~$\mu$m and a nominal minimal loss value of 0.495~dB/km at a wavelength of 1705~nm, and 0.6253 dB/km at 1550 nm for reference. The 3 dB/km bandwidth spans from 1225 nm to 2175 nm, which is a 950 nm range centered around 1750 nm, while the 1 dB/km bandwidth spans from 1450 nm to 2050 nm, which is a 600 nm range centered around 1750 nm. This indicates that our experiments are well within the optimal transmission range of the fiber. Theoretical investigations showed that effective coupling to the NANF fiber, up to $98\%$, can be achieved when the laser beam waist size is $70\%$ of the core diameter~\cite{Zuba2023Limits}. This was calculated to be 26 $\mu$m from the geometrical properties of the NANF as seen in the scanning electron microscope image shown in the inset of Fig.~\ref{fig:experimental_setup}. The beam diameter of the 1560 nm laser source was measured with a SWIR InGaAs camera, and a 40 mm CaF$_2$ plano-convex lens was used to achieve the desired beam size. 

The cutback method was used to measure the coupling losses and then the total transmittance through the 1-km long fiber was measured to establish the attenuation losses. By taking into account the coupling losses, the estimated attenuation values were 0.52 dB/km and 0.95 dB/km at 1560 nm and 2000 nm, respectively.  The attenuation loss measured for 2 $\mu$m is much lower than that of commercially available fused-silica fibers and is an improvement when compared to hollow-core photonic bandgap fibers (HC-PBGFs) with reported attenuation of 2.5 dB/km at 2~$\mu$m~\cite{Liu2015}. The measured attenuation is also close to the record attenuation value of 0.85 dB/km achieved in a five-tube NANF ~\cite{Zhang2022}.

The polarization extinction ratio (PER)  is a measure of how well a fiber maintains the polarization state of transmitted photons and is an important factor when it comes to measurements or systems that rely on light beams with the highest possible polarization purity \cite{stevens2002}.  We measured the PER for the NANF for a set of input polarization angles across the C-band with an InGaAsP tunable laser. To this end, a Glan-Taylor polarizer, a halfwave plate, and a thin film polarizer were used at the input of the fiber to set the polarization of the propagated light, with polarization optics repeating in reverse order at the output.  The polarizers were oriented to align with the horizontal axis of polarization and then fixed in place, while the halfwave plates are changed for state preparation and measurement.   The output light was focused on an InGaAs amplified detector connected to a lock-in amplifier. 

A squeezed vacuum state was generated through the nonlinear process of parametric down-conversion (PDC) at the wavelength of 2.09 $\mu$m. A schematic of the setup used is shown in Fig.~\ref{fig:experimental_setup}. To drive the nonlinear process, we used a ytterbium-based ultrashort pulse fiber laser (Chromacity Ltd.) with an average power of up to 2.5 W, a repetition rate of 80 MHz, a pulse duration of $\sim$127~fs, and a carrier wavelength of $\sim$1.045 $\mu$m. The pump beam was focused into a 1-mm-long periodically poled, magnesium-doped lithium niobate crystal to generate down-converted photons with type-0 phase-matched configuration (MgO-PPLN; Covesion Ltd.) with a poling period of 30.8 $\mu$m. Any remaining pump was filtered using a long-pass antireflection-coated germanium window blocking radiation $<1.85~\mu$m. 

The down-converted light is emitted over a broad angular spectrum. Thus, a 50-mm-focal-length gold-coated off-axis parabolic mirror was used to collect and collimate the PDC field before coupling to the NANF. A knife edge measurement was carried out to estimate the size of the down-converted beam. With a diameter of 6 mm (1/e$^2$), a CaF$_2$ lens of focal length $f=40$~mm was used to couple and propagate the PDC light into the hollow-core fiber. The radiation at the output was then collimated using a $f=40$~mm lens and then passed into the previously mentioned polarization analysis setup. Lastly, the squeezed vacuum was coupled to a 15-m-long single-mode fiber  (SM2000)  which was routed into a cryostat, where superconducting nanowire single-photon detectors (SNSPDs) were used for detection at the single-photon level. A 55-nm passband filter was used to select a portion of the down-converted field. The dispersion caused by the NANF fiber on the broadband radiation was also estimated by analyzing the arrival time of the down-converted pulses.

\section{Results}

\begin{figure}[h!]
    \centering
    \includegraphics[width=1\textwidth]{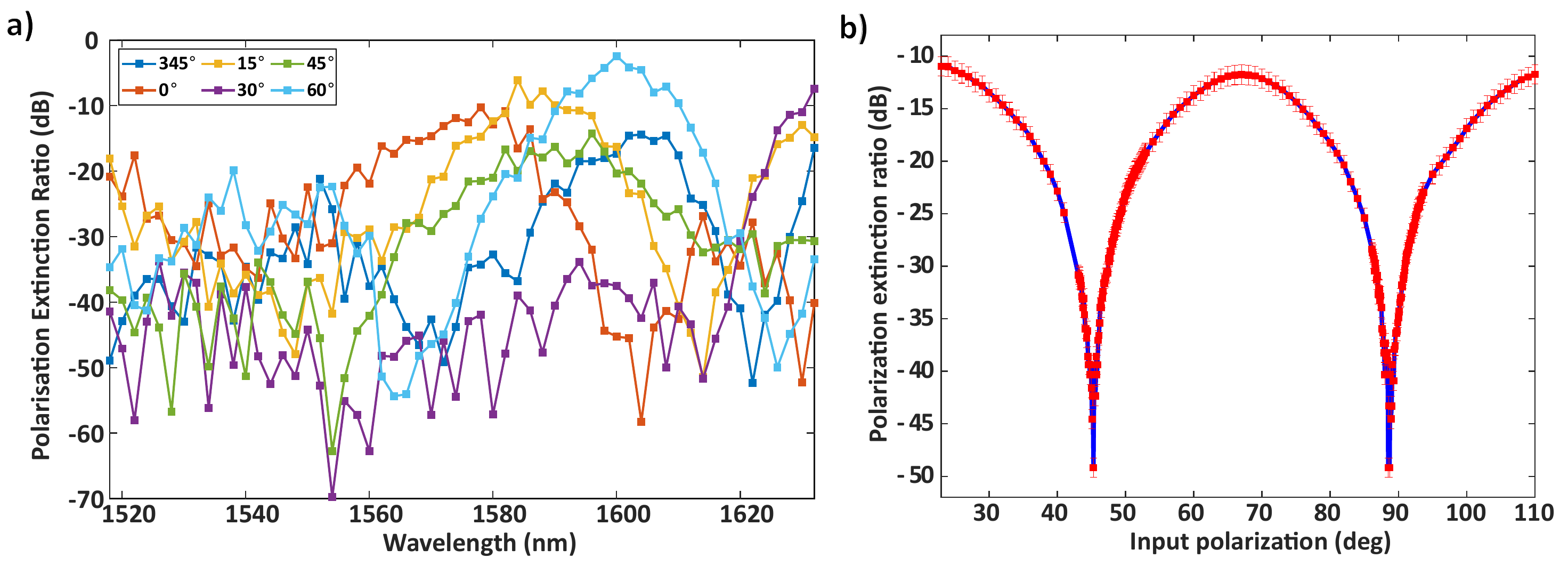}
    \caption{Measured polarization extinction ratio (PER) for C-band and 2 $\mu$m continuous-wave (CW) sources propagating through the NANF. a) PER measured across the C-band, with an inset showing the angle of the input half-wave plate, i.e., half the input polarization angle. b) PER as a function of input polarization angle at the 2 $\mu$m wavelength.}
    \label{fig:cbandPolPER}
\end{figure}

First, we investigated the effects of the fiber on CW light. This involved measuring the PER using the procedure outlined in the Methods section. Using a tunable laser source, we tested across the entire C-band of the telecom wavelength spectrum.  Figure 2 shows the variation of PER with different input polarization angles across a range of wavelengths.  The results in Fig.~\ref{fig:cbandPolPER} a) show an average optimal PER of -58 dB across various input polarization angles. Notably, at the wavelength of 1554 nm, the optimal PER value of -70 dB was achieved with an input polarization angle of 60$^\circ$ to the horizontal (which aligns with the plane of the fiber's coil), resulting from setting the HWP at 30$^\circ$. This is in close agreement with the state-of-the-art values obtained for NANF fibers designed for the 1550 nm waveband~\cite{taranta2020}. The observed fluctuations in PER could be due to a combination of factors. The principal axis of the fiber can vary with wavelength and may not be perfectly aligned along the entire spool length, leading to variations in birefringence and, consequently, affecting the PER. This misalignment causes the PER to change with the input polarization angle, introducing variability in the PER measurements (see also the supplementary information for Ref.~\cite{taranta2020}).

To investigate the 2-$\mu$m spectral region, we conducted a similar CW measurement. We utilized a Thulium-doped fiber laser, coupling its output to the NANF fiber to assess the PER with a horizontal input polarization. The findings, presented in Fig.~\ref{fig:cbandPolPER} b), indicate an optimal PER of -50 dB. For context, we note that standard single-mode silica fibers achieve a polarization purity of -30 dB at the 1550 nm wavelength\cite{sezerman1997accurate} which tends to worsen with increasing length and environmental stress. 
Exceptional polarisation purity in NANFs has been demonstrated to be robust  over various coil diameters (80 to 320~mm),  lengths (30~m to over 500~m) and against thermal stress at 1550 nm~\cite{taranta2020}, demonstrating that the polarization purity is not significantly affected by the winding and stress across various fibers and coil diameters. This points to greater feasibility of long-distance communication with NANF. We expect  this robustness to persist in the 2-$\mu$m spectral region.

It is well understood that the overall PER of a composite system is essentially limited by the component with the lowest PER, as derived from the multiplication of the Mueller matrices of the system components \cite{kraemer2018extinction}. In the 2-$\mu$m  waveband, the PER of the polarization optics without the NANF is approximately 55 dB. In the telecom C band, the PER of the polarization optics is approximately 80 dB.

\begin{figure}[h!]
    \centering
\includegraphics[width=0.85\textwidth]{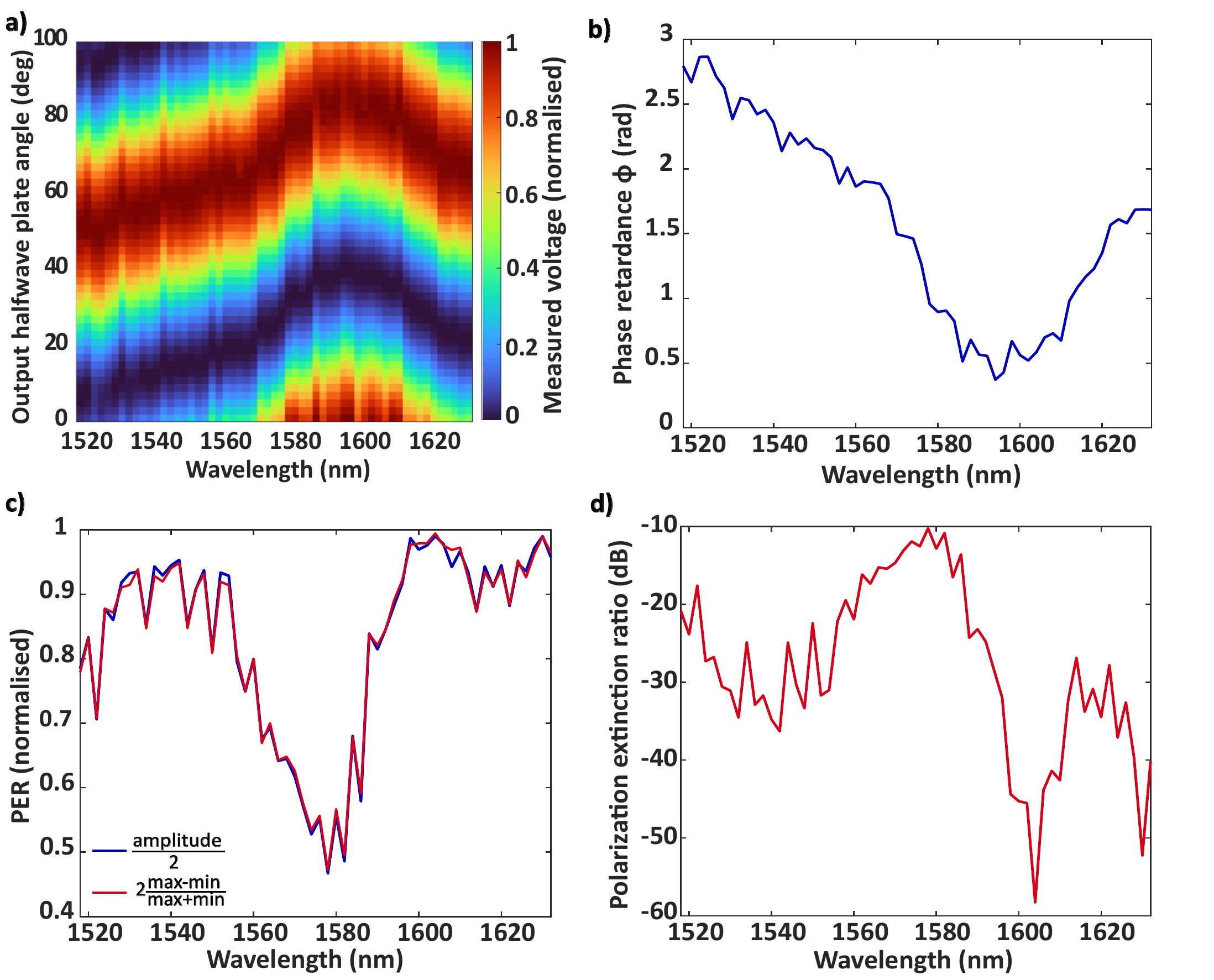}
    \caption{Linearly polarized light at C and L bands through the NANF. (a) Shows the self-normalized version of the signal amplitude obtained from the lock-in amplified photodiode detector for the various measurement half-wave plate angles at the output of the NANF. (b) Phase retardation (c) Normalized PER [as $2(\text{max}-\text{min})/(\text{max}+\text{min})$, where `min' and `max' refer to the minimum and maximum transmission values] obtained from fitting the voltage data to a cosine function. (d) PER in dB [as $10 \log_{10} \left( \text{min}/\text{max} \right)$]. The optimum value achieved with the input half-wave plate at  0$^\circ$ was -58~dB and occurred when the phase retardation was close to a minimum as it can be seen from (b).}
    \label{fig:pol_propagation_CL_band}
\end{figure}

Previous research has demonstrated a wavelength-dependent phase retardance, described by a Jones matrix model~\cite{taranta2020}. Here, we observe a similar wavelength dependence that results in varying power throughput for different wavelengths within the C-band spectrum, as shown in  Fig.~\ref{fig:pol_propagation_CL_band} (a).  
The results presented in Fig.~\ref{fig:pol_propagation_CL_band} (b)-(d)  show the case where the input half-wave plate angle was set to 0$^\circ$, corresponding to horizontal input polarization. 

\begin{figure}[h]
    \centering
\includegraphics[width=0.7\textwidth]{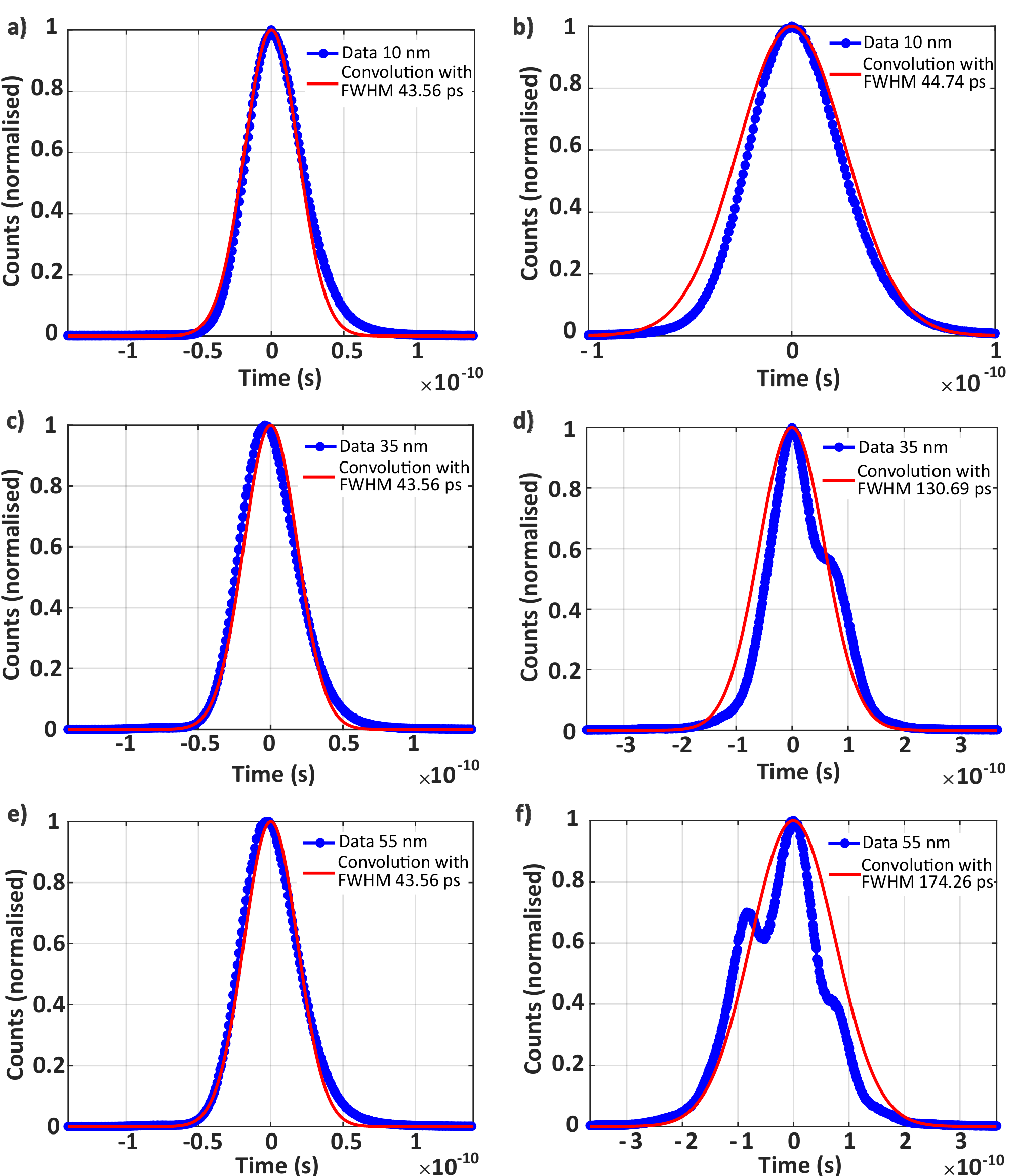}
    \caption{Dispersion introduced by the NANF. Subfigures (a) and (b) show the pulse after propagating through the 15-m-long SM2000 fiber and its elongated version from propagating through both the SM2000 fiber and the 1-km-long NANF, respectively, both with a 10 nm bandwidth input. Subfigures (c) and (d) illustrate the same for a 35 nm bandwidth input, and subfigures (e) and (f) for a 55 nm bandwidth input.}
    \label{fig:dispersion_NANF}
\end{figure}

In contrast to the CW sources that were used to characterize the PER of the NANF, the PDC field propagated through the fiber is broadband and pulsed, leading to effects like pulse dispersion introduced by the fiber. The dispersion introduced by the NANF will result in the spreading of a light pulse in time as it propagates down the NANF's length. To study this effect, the down-converted light was coupled into a 15-m-long single-mode SM2000 fiber which was routed to SNSPDs, and the photon arrival times were recorded using time-tagging electronics. The group delay dispersion (GDD) for different bandwidths of the pulsed down-converted radiation was obtained by using three different band-pass filters with Full Width at Half Maximum (FWHM) at 55, 35 and 10 nm.  A transform-limited Gaussian profile was assumed for each filter, with central wavelength at 2090 nm and FWHM given by the filter’s bandwidth. To quantify the temporal broadening, the duration of the pulses was measured using the PDC before propagation through the NANF and compared with detection at the output of the fiber (see 
Fig.~\ref{fig:dispersion_NANF}). The first column in Fig.~\ref{fig:dispersion_NANF} shows the measurement directly into the SMF2000 fiber, and the second column shows the measurement after the the light has traversed both the 1-km-long NANF fiber and the  15-m-long SM2000 fiber.

To analyze the broadening effect of the dispersion from the single-mode fiber on the broadband field, we convolved the transform-limited pulse with a Gaussian, which embodies the effects due to detector and electronic jitter. The FWHM of the Gaussian was then varied until the obtained convolution matched the experimental data. Similarly, a second Gaussian convolution was performed to match the experimental data obtained after propagating through the NANF, and a value for the FWHM of the broadened pulse was obtained. We observe that the width of the transform-limited pulse is predominantly determined by the jitter of our detection electronics, which is $\sim$4 ps, considerably larger than the 150-fs duration of the optical pulses used for characterization. Our method allowed us to ignore any dispersive effects introduced by components other than the NANF, such as the fibers outside and inside the cryostat housing the SNSPDs. Note that the pulse duration measured without propagation in the NANF is influenced by the detection jitter and the dispersion in the SM2000 fiber.

\begin{figure}[h]
    \centering
\includegraphics[width=0.5\textwidth]{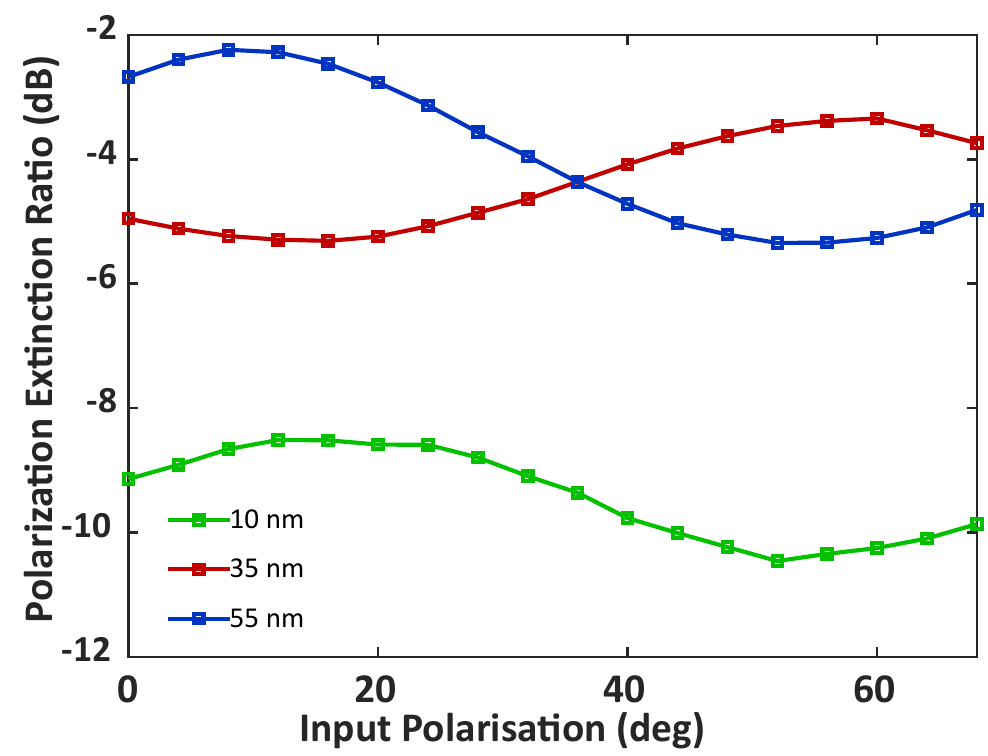}
    \caption{Polarization extinction ratio versus input polarization angle for vertically polarised  broadband photon pairs. The plot includes cases with the broadband photons generated by SPDC filtered using 10-, 35- and 55-nm spectral filters,  respectively. }
    \label{fig:PER_broadband}
\end{figure}

The setup used to perform PER measurements with the PDC field is shown in Fig.~\ref{fig:experimental_setup}. The PDC field was set to have linear polarization and then propagated through the NANF with light at the output coupled to the SNSPDs for detection. The PDC was passed through bandpass filters and the resulting PER values are plotted in Fig.~\ref{fig:PER_broadband}. The best PER was achieved when the broadband source was filtered using the 10-nm-bandwidth filter, which was the narrowest of the three.

\section{Discussion}
 \subsection{Potential for polarization-based quantum communication at 2-$\mu$m}
Further analyzing the data in Fig.~\ref{fig:PER_broadband} we gain insights into the suitability of the NANF as a channel in quantum-communication applications such as quantum key distribution (QKD).   In such cases, the quantum bit error rate (QBER) is a crucial parameter in assessing the security, efficiency and reliability of the protocol. It quantifies the rate at which errors occur during the transmission and measurement of quantum states. A high QBER can indicate potential eavesdropping or other vulnerabilities in the transmission. One of the factors that can negatively impact the QBER is the polarization extinction ratio (PER) of the fiber optic channel.  From the data, we can deduce the corresponding QBERs for both prepare-and-measure and entanglement-based protocols. 

{\bf \emph{BB84 Protocol.}} We consider the Bennett-Brassard 1984 (BB84) protocol, which stands as a commonly implemented prepare-and-measure QKD scheme~\cite{bennett1984,bennett1992}. Here, 4 quantum states are prepared by the sender and sent to the receiver through the quantum channel. Implementing it in terms of linear polarization bases, we can denote the quantum states as \(\{|H\rangle, |V\rangle, |\oplus\rangle, |\ominus\rangle\}\), where \(|H\rangle\) represents a horizontally polarized photon, \(|V\rangle\) represents a vertically polarized photon, \(|\oplus\rangle\) represents a photon polarized at +45° from the horizontal, and \(|\ominus\rangle\) represents a photon polarized at -45° from the horizontal. When a quantum state from this set is sent through the channel, it can be measured by the receiver using two different bases: the rectilinear/$Z$-basis (represented by$ \{|H\rangle, |V\rangle\}$), or the diagonal/$X$ basis (represented by \(\{|\oplus\rangle, |\ominus\rangle\}\)). 

Imperfections in the fiber or its alignment can lead to unwanted coupling between orthogonal polarizations degrading the PER, which in turn increases the QBER and could ultimately compromise the integrity of the quantum channel. 
The contribution to the QBER $\equiv \mathcal{E}$ from channel imperfections in this case can be expressed as 
\begin{equation}
\mathcal{E} = \frac{1}{4} \left( |\langle H_p|V_m\rangle|^2 + |\langle V_p|H_m\rangle|^2  + |\langle \oplus_p|\ominus_m\rangle|^2  +| \langle \ominus_p|\oplus_m\rangle|^2  \right),
  \label{eq:qber_overlap}
\end{equation}
where the subscripts $p,m$ refer to the prepared and measured states, respectively. We note that each term in the Eq.~\ref{eq:qber_overlap} is a conditional probability which can be deduced from the PER for input polarization states $|H_p\rangle, |V_p\rangle, |\oplus_p\rangle, |\ominus_p\rangle$. 

The results shown in Fig.~\ref{fig:PER_broadband} correspond to QBERs of $10.97\%$,  $33.91\%$, and $36.31\%$ for the 2-$\mu$m broadband photon pairs using 10-, 35- and 55-nm spectral filters, respectively.  These QBERs were determined using an optimised choice of bases for the input polarisation states, where the state $|H_p\rangle$ corresponded to input angles $23^\circ$, $0^\circ$ and $23^\circ$ (in Fig.~\ref{fig:PER_broadband}) for  measurements with 10-, 35- and 55-nm spectral filters, respectively.  
 11\% is considered the upper limit for secure implementation of BB84 QKD~\cite{gottesman2004security}. A QBER exceeding this value is an indication that the ability to distill a secure key is compromised. This suggests that with the fiber under study, polarization-based BB84 QKD is promising at bandwidths below  $\approx$10 nm---a limit imposed by considering only the polarization purity of the fiber. We emphasise that these results do not reflect the ultimate performance limit of the NANF technology and future work to further optimise the performance in the 2-$\mu$m region is expected to result in considerably improved values. 

{\bf\emph{BBM92 Protocol.}} Entanglement-based protocols such as  BBM92~\cite{bennett1992} rely on coincidence measurements. This naturally gives them synchronizability and thus operability in CW mode~\cite{yin2020entanglement}, for which the fiber shows a much better performance as seen in Fig.~\ref{fig:cbandPolPER} b). In BBM92, two parties share a source that emits entangled photon pairs, e.g.,  in the Bell state 
 \begin{equation}
|\Phi^+\rangle = \frac{1}{\sqrt{2}} \left( |HH\rangle + |VV\rangle \right),
\end{equation}
where \(|HH\rangle\) and \(|VV\rangle\) are the states with both photons having horizontal and vertical polarizations respectively, while \(|HV\rangle\) and \(|VH\rangle\) represent states with one photon horizontally polarized and the other vertically polarized, and vice versa. The receiver for each party can choose to measure the incoming photons in either the $Z$ basis, or the $X$ basis. The security and efficiency of BBM92 are attributed to the utilization of entanglement, which guarantees the correlations observed between the two communicating parties. 
In this case, the secure key rate is given as\cite{PhysRevLett.90.057902,yin2020entanglement}

 \begin{equation}
R_{Z/X} \geq Q_{Z/X}\left[1- H\left(\mathcal{E}_Z\right)-H\left(\mathcal{E}_X\right)\right],
\end{equation}
where $Q_{Z/X}$ signifies the sifted key rate (per detected pair) corresponding to when both parties choose the same basis (${Z}$ or ${X}$, respectively).  $\mathcal{E}_Z$ and $\mathcal{E}_X$ denote the QBER values in the $Z$ and $X$ bases, respectively.  $H(\mathcal{E})$ is the binary entropy function defined as  $H(\mathcal{E})=-\mathcal{E}\log _2 \mathcal{E}-(1-\mathcal{E}) \log _2(1-\mathcal{E})$.  The overall asymptotic secret key rate is then given by $R_{\mathrm{A}}=R_Z+R_X$. 

The limiting asymptotic key rate corresponding to the PER measured using the broadband SPDC-generated photon pairs is found to be positive only for the 10-nm-bandwidth case (where QBER is optimised with $|H_p\rangle$ chosen as input angle $0^\circ$), giving $R_A=0.06$~bits as the value limited by only the PER of the fiber. For the CW case (0.2-nm linewidth at 2 $\mu$m), the PER measurements shown in Fig.~\ref{fig:cbandPolPER} b) correspond to a QBER of $2.7\%$ (optimized with $|H_p\rangle$ chosen as input angle $33^\circ$) and BBM92 asymptotic secret key rates of $0.68$~bits/pair of maximally entangled photons detected, again, limited by the PER of the NANF fiber.

\section{Conclusion}

In conclusion, this study has demonstrated the potential  and current limitations of exploiting the low-loss wavelength window around 2-$\mu$m for quantum communication using a specialized hollow-core fiber. 
	Moreover, the characterization of the fiber's dispersion properties reveals that it supports broadband operation, albeit limited by dispersion. The effects of dispersion can be mitigated using spectral filtering to achieve an uncompromised performance, crucial for multiplexing in quantum networks---a concept which is gaining more attention in recent advancements in free-space QKD where, e.g., spatial-multiplexing and time-division techniques are leveraged for improved throughput~\cite{TelloCastillo:22}.  
	
Analyzing the polarization characteristics, we find that the polarization extinction ratio (PER) significantly influences the quantum bit error rate (QBER) and secure key rates, which are a crucial parameters for quantum-communication protocols. For the BB84 protocol, it is apparent that the quality of the polarization-based quantum key distribution (QKD) is highly sensitive to the spectral bandwidth of the photons. With a QBER acceptable for secure quantum communication only at bandwidths approximately below 10 nm, this places a stringent limitation on the operability of a polarization-based BB84 protocol using this fiber, determined primarily by the polarization purity that the fiber can maintain. On the other hand, the BBM92 protocol, which utilizes entangled photons and allows for synchronization in a continuous-wave mode, seems more promising for implementation of polarization-based QKD with this fiber. The PER measurements and the subsequent analysis indicate that the fiber could sustain a robustly positive asymptotic secret key rate, thereby reinforcing the potential of the 2-$\mu$m window for both prepare-and-measure and entanglement-based quantum-communication protocols. We note that fibers are typically not considered ideal for polarization-based QKD due to inherent challenges, giving preference for using the temporal degree of freedom, such as time-bin encoding which benefits from low-latency propagation in air. However, the advent of NANF fibers also opens the possibility of not excluding polarization as a viable degree of freedom for long distance quantum communication. These results in the  2-$\mu$m window also further open up the bandwith capabilities of HCFs. Indeed, recent research~\cite{Nasti:22} has demonstrated the capability of NANFs to transmit single-photon and classical  optical channels separated by hundreds of nanometers with  a remarkable 30~dB reduction in background noise compared to using closer wavelengths. The use of HCFs in this context represents a promising advancement in the development of practical and scalable quantum networks that can coexist with classical communication infrastructure.

Our findings highlight the critical parameters that must be optimized for the development of reliable and secure quantum-communication systems. Furthermore, the results point to further research into the deployment of such fibers in a broader range of quantum technologies, possibly opening new avenues for quantum communication at higher data rates, over longer distances, and under more practical conditions.

\begin{backmatter}
\bmsection{Funding}
D.F. is supported by the Royal Academy of Engineering under the Chairs in Emerging Technologies scheme and the Engineering and UK Physical Sciences Research Council (EPSRC) project Quantic
(EP/T00097X/1). M.C. acknowledges support from the UK Research and Innovation (UKRI) and the EPSRC Fellowship (“In-Tempo” EP/S001573/1).  A.C.D. acknowledges support from the EPSRC Impact Acceleration Account (EP/R511705/1).

\bmsection{Acknowledgments}
We gratefully acknowledge the Optoelectronics Research Center (ORC) for the loan of the fiber used in this work. 

\bmsection{Disclosures}
The Authors declare no conflicts of interest.

\bmsection{Data Availability}
All the data supporting the conclusions reported in this manuscript are available online [\href{http://dx.doi.org/10.5525/gla.researchdata.1772}{http://dx.doi.org/10.5525/gla.researchdata.1772}]. Additional data related to this paper may be requested from the authors.  

\end{backmatter}



\end{document}